\documentclass[12pt]{iopart}

\usepackage{epsfig}
\usepackage{iopams}

\newcommand{\text}[1]{\textrm{#1}}
\newcommand{\bm}{\boldsymbol}
\newcommand{\eqref}[1]{(\ref{#1})}

\newenvironment{subequations}{\numparts}{\endnumparts}

\newcommand{\bo}[1]{\mathbf{#1}}
\newcommand{\order}{\Or\,}
\newcommand{\mc}[1]{\mathcal{#1}}
\newcommand{\wt}[1]{\widetilde{#1}}
\newcommand{\ba}[1]{\overline{#1}}
\newcommand{\op}[1]{\widehat{#1}}
\newcommand{\dagop}[1]{\widehat{#1}^{\dagger}}
\newcommand{\var}[1]{\text{var}\left[{#1}\right]}
\newcommand{\re}[1]{\text{Re}\left[{#1}\right]}

\begin{document} 

\title[First-principles quantum dynamics in interacting Bose gases I]
{First-principles quantum dynamics in interacting Bose gases I:
 The positive P representation}

\author{P Deuar and P D Drummond\footnote{www.physics.uq.edu.au/BEC}}

\address{Australian Centre for Quantum Atom Optics,
    The University of Queensland, Brisbane, Australia}

\eads{\mailto{deuar@physics.uq.edu.au}, \mailto{drummond@physics.uq.edu.au}}


\begin{abstract}
  The performance of the positive P  phase-space representation for exact many-body quantum dynamics is investigated. Gases of interacting bosons are considered, where the full quantum equations to simulate are of a Gross-Pitaevskii form with added Gaussian noise. This method gives tractable simulations of many-body systems because the number of variables scales linearly with the spatial lattice size. 
An expression for the useful simulation time is obtained, and checked in numerical simulations. The dynamics of first-, second- and third-order spatial correlations are calculated for a uniform interacting 1D Bose gas subjected to a change in scattering length. Propagation of correlations is seen. A comparison is made to other recent methods.
The positive P method is particularly well suited to open systems as no conservation laws are hard-wired into the calculation. It also differs from most other recent approaches in that there is no truncation of any kind.
\end{abstract}

\submitto{\JPA}


\section{Introduction}
\label{INTRO}

Simulation of exact many-body quantum dynamics is a highly non-trivial task even in the simplest models. However, such a capability is highly desirable for studies of the quantum behaviour of mesoscopic systems. 
  First principles methods are particularly useful when there are several length, time, or energy scales  of similar size, which makes simplifying approximations inaccurate.
It is well known, however, that a calculation using an orthogonal basis
 is intractable because the size of the Hilbert space needed to accurately describe the state grows exponentially with the number of subsystems. Perturbation theory is often not useful either, owing to questions of convergence. Other commonly used techniques -- like the mean-field approximation -- rely on  factorizations of operator products, with unknown regimes of validity.

A promising approach is to simulate stochastic equations derived from  phase-space distribution methods such as the positive P\cite{positiveP1,positiveP2} and gauge P\cite{stochasticgauges} representations, or representations based on other separable basis sets. 
In such methods the quantum state is written as a probability distribution of off-diagonal system configurations, which are separable between local subsystems. For the case of the positive P representation, these  subsystems are the spatial lattice points. The number of variables needed to specify a single configuration grows only \textit{linearly} with the system size  because they are separable. Correlations between subsystems are contained in the details of the distribution, which is stochastically sampled. 
This very mild growth of the number of required variables is the reason that such methods lead to tractable many-body simulations. Recent work has shown that  any model with two-body interactions can be expressed by such a distribution of separable system configurations\cite{separabledistributions}.

In this paper we develop reliable ways of estimating the feasible simulation time, which is limited in practise by the
growth of the statistical error. This allows an assessment of simulation parameters 
 before a potentially lengthy calculation is started. We apply our method to single and coupled anharmonic oscillators,
 and compare analytic and numerical results for the simulation time. We also investigate the quantum
dynamics of a one-dimensional interacting Bose gas. This calculation shows a unique and interesting behaviour under
conditions of a sudden switch in the interaction strength. We observe correlation waves in which an otherwise
homogeneous quantum gas develops wave-like structures only found in the relative coordinates.

 This is a particularly timely issue in view of recent developments in fibre optics and ultra-cold atomic Bose gases. It
 is now possible to experimentally investigate large numbers of interacting bosons, under conditions where quantum coherence plays an important role. As an example, quadrature squeezing in quantum solitons was predicted using these techniques, and observed experimentally. Ultra-cold atomic Bose gas dynamics has been observed in a number of geometries. In these experiments, it is possible to modify external potentials in order to drive the quantum gas into a state very far from either the ground state or any thermal equilibrium state. More recently, experiments have demonstrated the possibility
 of dynamically varying the  inter-particle potential.

The main body of the paper is organized as follows: Firstly the positive P method is summarized in Section~\ref{METHOD}, and briefly compared to other approaches in Section~\ref{HIST}. Subsequently 
in Sections~\ref{PPTIME} and ~\ref{MMODE},  the estimates of useful integration times are  obtained, and their implications considered. 
These estimates are verified by numerical simulations of single-mode  and multi-mode systems, in Sections~\ref{PPEMPIRICAL} and ~\ref{MEX}, respectively. 
As an example, in the latter Section, the dynamics of spatial correlation functions in a 1D uniform gas 
are investigated. We consider the effects of an experiment in which the inter-particle potential is suddenly changed. This
is potentially observable using the technique of a Feshbach resonance in an external magentic field. 

In a subsequent paper\cite{paperB}, we consider the more sophisticated stochastic gauge method for these types of simulations.

\section{Model}
\label{METHOD}

\label{SYSTEM}
\subsection{The lattice model}
\label{LATTICEMODEL}
We consider a dilute interacting Bose gas. In the continuum, the second-quantized Hamiltonian with kinetic terms, two-body interactions with interparticle potential $U(\bo{r})$, and external potential $V_{\rm ext}(\bo{x})$ is
\begin{eqnarray}\eqalign{
\op{H}  =\int &\left[\frac{\hbar ^{2}}{2m}{\bm \nabla }\dagop{\Psi }({\bo{x}}){\bm \nabla }\op{\Psi }({\bo{x}})+V_{\rm ext}({\bo{x}})\dagop{\Psi}({\bo{x}})\op{\Psi }({\bo{x}})\right.\\
 & \left.+\frac{U(\bo{x}-\bo{y})}{2}\dagop{\Psi }(\bo{x})\dagop{\Psi }({\bo{y}})\op{\Psi }({\bo{x}})\op{\Psi }({\bo{y}})\right]d^3{\bo{x}}.\label{model}
}\end{eqnarray}
 Here $\op{\Psi }({\bo{x}})$ is a boson field operator at ${\bo{x}}$ and
$m$ is the boson mass.

In practice,  (\ref{model}) is replaced by a 
lattice Hamiltonian, which contains all the essential
features provided the lattice spacing is sufficiently small. 
For a rarefied gas of the kind occurring in contemporary BEC experiments, $s$-wave scattering dominates\cite{interactionassumptions}, and the $s$-wave scattering length $a_s$ is much smaller than all other relevant length scales. If the lattice spacing is also much larger than $a_s$, then the two-body scattering is well 
described by an interaction local at each lattice point.

Let us label the spatial dimensions by $d=1,\dots,D$, and label lattice points by the vectors $\bo{n}=(n_1,\dots,n_D)$.
For lattice spacings $\Delta {\rm x}_d$, the spatial coordinates of the lattice points are $\bo{x}_{\bo{n}}= (n_1\Delta {\rm x}_1, \dots, n_D\Delta{\rm x}_D)$. 
The volume per lattice point is $\Delta V=\prod_d\Delta {\rm x}_d$. We also define the lattice
annihilation operators $\op{a}_{\bo{n}}\approx\sqrt{\Delta V}\,\op{\Psi }(\bo{x}_{\bo{n}})$, 
which obey the usual boson commutation
relations of $[\op{a}_{\bo{n}},\dagop{a}_{\bo{m}}]=\delta _{\bo{n}\bo{m}}$. 
With these definitions, one obtains:
\begin{equation}
\op{H}=\hbar \left[\sum _{\bo{n}\bo{m}}\omega _{\bo{n} \bo{m}}\,\dagop{a}_{\bo{n}}\op{a}_{\bo{m}}+\frac12\sum _{\bo{n}}\kappa \op{a}^{\dagger 2}\op{a}^2\right]\, .\label{Hamiltonian}\end{equation}
In this normally ordered Hamiltonian, the  frequency terms
$\omega _{\bo{n} \bo{m}}=\omega^*_{\bo{m} \bo{n}} $
come from the kinetic energy and external potential. They produce a local particle-number-dependent energy, and linear coupling to other sites, the latter arising only from the kinetic processes. The nonlinearity due to local particle-particle interactions is of strength
\begin{equation}
\kappa= \frac{g}{\hbar\,\Delta V},
\end{equation}
with the standard coupling value\cite{interactionassumptions} being $g=g_{\rm 3D}=4\pi a_s\hbar^2/m$ in 3D, and 
$g=g_{\rm 3D}/\sigma$ in 2D and 1D, where $\sigma$ is the effective thickness or cross-section of the collapsed dimensions.

When interaction with the environment is Markovian (i.e. no feedback), the evolution of the density matrix $\op{\rho}$ can be written as a master equation in Linblad form
\begin{equation}\label{master}
\frac{\partial\op{\rho}}{\partial t} = \frac{1}{i\hbar}\left[\op{H},\op{\rho}\right] +\frac{1}{2}\sum_j\left[2\op{L}_j\op{\rho}\dagop{L}_j-\dagop{L}_j\op{L}_j\op{\rho} - \op{\rho}\dagop{L}_j\op{L}_j \right].
\end{equation}
  For example, single-particle losses at rate $\gamma_{\bo{n}}$ (at $\bo{x}_{\bo{n}}$) to a \mbox{$T=0$} heat bath are described by $\op{L}_{\bo{n}}=\op{a}_{\bo{n}}\sqrt{\gamma_{\bo{n}}}$.

\subsection{Positive P representation}
\label{GAUGEP}
The positive P representation was developed in \cite{positiveP1,positiveP2}. Here we summarize the issues relevant to the dynamics of the model (\ref{model}), and present the stochastic equations to simulate.

For a lattice with $M$ points, the density matrix is expanded as
\begin{equation}\label{gaugerep}
\op{\rho} = \int P_+(\bm \alpha, \bm \beta)\, \op{\Lambda}(\bm \alpha, \bm \beta)\, d^{2M}\!\bm \alpha\,d^{2M}\!\bm \beta,
\end{equation}
where $\op{\Lambda}$ is an off-diagonal operator kernel, separable between the $M$ lattice point subsystems. We use a coherent-state basis so that
\begin{equation}\label{kernel}
\op{\Lambda}(\bm \alpha, \bm \beta) = 
||\,\bm\alpha\rangle\langle \bm\beta^*||\exp[-\bm\alpha\cdot\bm\beta],
\end{equation}
in terms of Bargmann coherent states with complex amplitudes $\bm\alpha=(\dots,\alpha_{\bo{n}},\dots)$:
\begin{equation}
||\,\bm\alpha\rangle = \otimes_{\bo{n}}\exp\left[ \alpha_{\bo{n}} \dagop{a}_{\bo{n}} \right]|\,0\rangle.
\end{equation}
The gauge P representation mentioned previously\cite{stochasticgauges} includes an extra  global weight in the kernel, which allows useful modifications of the final stochastic equations\cite{removalofboundaryterms,stochasticgauges,paperB,inprep}.

 It has been shown that all density matrices can be represented by a positive real $P_+$\cite{positiveP2}. The expansion (\ref{gaugerep}) then becomes a probability distribution of the $\op{\Lambda}$, or equivalently of the variables $\bm\alpha$, $\bm\beta$. A constructive expression for  $P_+(\op{\rho})$  is  given by expression (3.7) in reference~\cite{positiveP2}, although more compact distributions may exist. In particular, a coherent state $|\bm\alpha_o\rangle$ will simply have
\begin{equation}
P_+ = \delta^{2M}(\bm\alpha-\bm\alpha_o)\,\delta^{2M}(\bm\beta-\bm\alpha_o^*).
\end{equation}

Using the identities 
\begin{subequations}\label{id}
\begin{eqnarray}
\op{a}_{\bo{n}}\op{\Lambda} &=& \alpha_{\bo{n}}\op{\Lambda},\label{ida}\\
\dagop{a}_{\bo{n}}\op{\Lambda} &=& \left[\beta_{\bo{n}}+\frac{\partial}{\partial \alpha_{\bo{n}}}\right] \op{\Lambda},\label{idad},
\end{eqnarray}\end{subequations}
the master equation (\ref{master}) in $\op{\rho}$ can be shown to be equivalent to a Fokker Planck equation in $P_+$, and then to stochastic equations in the $\alpha_{\bo{n}}$ and $\bm\beta_{\bo{n}}$.  The standard method is described in reference~\cite{gardiner}. The correspondence is in the sense that appropriate stochastic averages of $\alpha_{\bo{n}}$ and $\beta_{\bo{n}}$ correspond to quantum expectation values in the limit when the number of trajectories $\mc{S}\to\infty$. In particular one finds that\cite{removalofboundaryterms}
\begin{eqnarray}\label{moments}
\left\langle\prod_{jk} \dagop{a}_{\bo{n}_j}\op{a}_{\bo{n}_k}\right\rangle = 
\lim_{\mc{S}\to\infty}\left\langle \prod_{jk} \beta_{\bo{n}_j}\alpha_{\bo{n}_k}  + 
\prod_{jk} \beta^*_{\bo{n}_k}\alpha^*_{\bo{n}_j}
\right\rangle_{\rm s}.
\end{eqnarray}
Any observable can be written as  a linear combination of terms \eqref{moments}. 
We will use the notation $\langle\cdot\rangle_{\rm s}$ to distinguish stochastic averages of random variables from quantum expectations $\langle\cdot\rangle$. 

The resulting $2M$ Ito complex stochastic equations  are found using  methods described in  Refs.~\cite{positiveP2,gardiner}. For the model (\ref{Hamiltonian}) obeying the master equation (\ref{master}) with coupling to a zero temperature heat bath, they are of the Gross-Pitaevskii (GP) form, plus noise:
\begin{eqnarray}\eqalign{\label{itoequations}
d\alpha_{\bo{n}} =& [-i\sum_{\bo{m}} \omega_{\bo{n}\bo{m}} \alpha_{\bo{m}} -i\kappa n_{\bo{n}}\alpha_{\bo{n}} -\gamma_{\bo{n}}\alpha_{\bo{n}}/2]dt + \sum_k B^{(\alpha)}_{\bo{n}k}\,dW_k,\\
d\beta_{\bo{n}} =& [\ \ i\sum_{\bo{m}} \omega_{\bo{n}\bo{m}}^* \beta_{\bo{m}} +i\kappa n_{\bo{n}}\beta_{\bo{n}} -\gamma_{\bo{n}}\beta_{\bo{n}}/2]dt + \sum_k B^{(\beta)}_{\bo{n}k}\,dW_k.
}\end{eqnarray}
Here,  $n_{\bo{n}} = \alpha_{\bo{n}}\beta_{\bo{n}}$, and there are $M'\ge2M$ labels $k$ to sum over. The $dW_k$ are independent Wiener increments $\langle dW_j(t)\,dW_k(s)\rangle_{\rm s}=\delta_{jk}\delta(t-s)dt^2$. In practice, these can be realized at each time step $\Delta t$ by independent real Gaussian noises of mean zero and  variance $\Delta t$.
The elements of the $M\times M'$ \textit{noise matrices} $B$ must satisfy
\begin{eqnarray}\eqalign{\label{BB=D}
  \sum_k B^{(\alpha)}_{\bo{n}k} B^{(\alpha)}_{\bo{m}k} =& -i\kappa\alpha_{\bo{n}}^2\delta_{\bo{n}\bo{m}}\label{BB=D1}\\
  \sum_k B^{(\beta)}_{\bo{n}k} B^{(\beta)}_{\bo{m}k} =& i\kappa\beta_{\bo{n}}^2\delta_{\bo{n}\bo{m}}\label{BB=D2},\\
  \sum_k B^{(\alpha)}_{\bo{n}k} B^{(\beta)}_{\bo{m}k} =& 0.
}\end{eqnarray}
These do not actually specify the elements of $B$ completely. The usefulness of these degrees of freedom are considered in detail in Refs.~\cite{stochasticgauges,paperB,inprep}, but here we will just use the simplest choices:
\begin{eqnarray}\eqalign{\label{stdpp}
B^{(\alpha)}_{\bo{n}_jk} =& i\sqrt{i\kappa}\alpha_{\bo{n}_j}\delta_{j,k},\\
B^{(\beta)}_{\bo{n}_jk} =& \sqrt{i\kappa}\beta_{\bo{n}_j}\delta_{(j+M),k}
}\end{eqnarray}
(with $k=1,\dots,2M$).
The indices $j=1,\dots,M$ label the lattice points. 

The simulation strategy is (briefly): \begin{enumerate}
\item Sample a trajectory according to the known initial condition $P_+(0)=P_+(\,\op{\rho}(0)\,)$ 
\item Evolve according to the stochastic equations (\ref{itoequations}), calculating moments of interest, and recording.
\item Repeat for $\mc{S}\gg1$ independent trajectories and average.
\end{enumerate}

\section{Comparison with other dynamical methods}
\label{HIST}

Several positive P and gauge P simulations of the models of Section~\ref{LATTICEMODEL}, (or very similar) have already appeared:
\begin{itemize}
\item Quantum optical soliton propagation in nonlinear Kerr media\cite{kerrsolitons}, which obeys a Hamiltonian formally very similar to (\ref{model}). 
\item  Dynamics of evaporative cooling and incipient condensation of an interacting Bose gas\cite{evaporativecooling,samperrorinbec}. 
\item   Thermal and dynamical behaviour of spatial correlations in one-dimensional interacting Bose gases\cite{1dbosethermo,inprep} using the stochastic gauge representation\cite{removalofboundaryterms,stochasticgauges}. 
 \end{itemize} 

Some of the other many-mode dynamics simulation methods proposed more recently include (See Table~\ref{TABLEmethods} for a visual comparison):

  The \textbf{stochastic wavefunction} method of Carusotto, Castin and Dalibard\cite{carusottodynamix,carusottothermo} shares many common features with the gauge P method\cite{inprep}. It is also based on a phase-space distribution over separable operators with a global weight, and can be regarded formally as a particular choice of gauge and basis set. In this case the separable ``subsystems'' are chosen to be exactly $N$ identical particles, rather than the $M$ spatial lattice points of the gauge P/positive P approach. Hence, sampled variables specify a orbital occupied by these $N$ particles, rather than the conditions at each lattice point. This explicitly imposes both a precise particle number and particle conservation on the model, and so makes the method directly applicable only to closed models where processes such as random losses, outcoupling, or gain are absent. Where both stochastic gauge and stochastic wavefunction methods can be applied, the useful simulation time is similar\cite{paperB}.

  A different approach entirely is based on extensions of the \textbf{density matrix renormalization group} (DMRG) method\cite{DMRG-Caltech,DMRG-Garching}. This method is based on the concept that an $N$-subsystem state is equivalent to a  construction called a matrix product state (MPS) in a larger Hilbert space.  This involves 
 a further $N$ ``ancillary'' systems entangled with the original subsystems. In cases where entanglement between the original subsystems is small or short-range (such as spin chains with nearest-neighbour interactions), a good approximation to the full state can be obtained by truncating the amount of entanglement present in the MPS construction. These truncated states can be stored efficiently with a number of variables polynomial in $N$, due to the way the MPS states are constructed.  This can lead to tractable \textit{deterministic} calculations even with significant $N$. 
 For longer-range coupling in 2D and 3D, there are indications based on the entropy of bipartite entanglement that 
the amount of entanglement is expected to grow too fast for tractable accurate simulations with this method\cite{DMRG-bipartiteentanglement}.

\begin{table}
\caption{ Some general features of several methods for simulation of many-body dynamics. $M$ is the number of lattice points, $N$ the number of particles. $d$ is the Hilbert space dimension of the subsystem at each lattice point, $D$ the dimension of each corresponding ancillary state after possible truncation of entanglement in the DMRG-based methods.
\label{TABLEmethods}
}
\begin{tabular}{@{}lccc}\br
				&stochastic gauge		& stochastic		& DMRG-based 		\\
				& positive P		& wavefunction		& methods		\\
\mr
subsystems 			& lattice points 	& particles		& $d-$level systems	\\
stochastic 			& yes			& yes			& no			\\
hard-wired conservation laws	& none			& particle number 	& $d$-level systems	\\
open systems 			& yes			& no			& yes			\\
truncation			& none			& none			& entanglement		\\
tractable coupling		& wide variety		& wide variety		& short-range or 1D	\\
memory requirements 		& $\propto M$		& $\propto M$		& $\propto MdD$		\\
Hilbert space dimension		& $\infty^M$		& $M^N$			& $d^M$			\\
references			& \cite{positiveP1,positiveP2,removalofboundaryterms,stochasticgauges,1dbosethermo,inprep}
							&\cite{carusottodynamix,carusottothermo}
										& \cite{DMRG-Caltech,DMRG-Garching,DMRG-bipartiteentanglement}
													\\\br
\end{tabular}	
\end{table}

While each approach has its own merits, for large and/or 2D, 3D systems the stochastic phase-space methods appear to be the most efficient, perhaps apart from special cases. There, for the price of limited precision, one can obtain an un-truncated first-principles simulation whose demands on resources and time scale only linearly with size. For open systems, the methods based on lattice point subsystems (such as the positive P) can be applied in a straightforward manner, as they explicitly allow a varying particle number. In contrast, attempts to use orbital-based states as in the stochastic wavefunction lead at best to additional complication.

\section{Single-mode simulation times}
\label{PPTIME}

\subsection{Exactly solvable single-mode model}
\label{PPTIME1MODE}
The single-mode case is of special interest as it is exactly solvable, while still describing an
interacting many-body quantum system.  It already contains the essential features  that make the distribution of trajectories broaden, thus limiting the useful simulation time. 

To simplify the notation, the mode labels $\bo{n}=(0,\dots,0)$ will be omitted when referring to the single-mode system. Furthermore, we move to an interaction picture where the harmonic oscillator evolution due to the $\hbar\omega\dagop{a}\op{a}$ term in the Hamiltonian is implicitly contained within the Heisenberg evolution of the operators. Then, this ``anharmonic oscillator'' model has 
\begin{equation}\label{1modeH}
\op{H} = \frac {\hbar\kappa}{2}\,\op{a}^{\dagger 2}\op{a}^2,
\end{equation}
and the  equations to simulate are
\begin{eqnarray}\eqalign{\label{1modeppequations}
d\alpha = \alpha&[-i\kappa n\,dt -\gamma\,dt/2+i\sqrt{i\kappa}\,dW_1],\\
d\beta = \beta&[\ \ i\kappa n\,dt -\gamma\,dt/2+\sqrt{i\kappa}\,dW_2\ ],
}\end{eqnarray}
with $n=\alpha\beta$. Note that by (\ref{moments}), the mode occupation $\langle\dagop{a}\op{a}\rangle$ is $\langle\re{n}\rangle_{\rm s}$.

These equations can be formally solved using the rules of Ito calculus as 	
\begin{subequations}\label{formalpp}\begin{eqnarray}
\log n(t) &=& \log n(0) -\gamma t +\sqrt{i\kappa}\left[ \zeta_2(t) + i\zeta_1(t) \right],\qquad\label{nformalpp}\\
\log\alpha(t) &=& \log\alpha(0) +(i\kappa-\gamma)t/2 +i\sqrt{i\kappa}\,\zeta_1(t) -i\kappa\int_0^t n(s)\,ds, 
\end{eqnarray}
where
\begin{equation}\label{zetadef}
\zeta_j(t) = \int_0^{W(t)} dW
\end{equation}\end{subequations}
are Gaussian-distributed random variables of mean zero. However, they are not time-independent:
\begin{equation}\label{zetamean}
\langle \zeta_j(t) \zeta_k(s) \rangle_{\rm s} = \delta_{jk}\,\text{min}\left[ t, s\right].
\end{equation}
In what follows it will be convenient to consider the evolution of an off-diagonal coherent-state kernel 
\begin{equation}\label{offkernel}
\op{\Lambda}_0=\op{\Lambda}(\alpha_0,\beta_0),
\end{equation}
 with complex ``particle number'' $n_0=\alpha_0\beta_0$. This is because for any general initial state, each sampled trajectory will start out as an $\op{\Lambda}_0$ with some coherent amplitudes.

With initial condition $\op{\Lambda}_0$, analytic expressions for observables can be readily obtained. In particular, 
we will calculate the first-order time-correlation function  
 \begin{equation}\label{G1tdef}
G^{(1)}(0,t)=\beta_0\langle\op{a}\rangle = \alpha_0^*\langle\dagop{a}\rangle^*
 ,\end{equation}
 which contains phase coherence information. Normalizing by $\text{Tr}[\op{\Lambda}_0]$, 
\begin{equation}\label{G1exact}
G^{(1)}(0,t) = n_0\,e^{-\gamma t/2}\,
 \exp{\left\{\frac{n_0}{1-i\gamma/\kappa}\left(e^{-i\kappa t-\gamma t}-1\right)\right\}}.
\end{equation}

  When the damping is negligible, $n_0$ real, and the number of particles is $n_0\gg 1$, one sees that the initial phase oscillation period is
\begin{equation}
t_{\text{osc}}=\frac{1}{\kappa}\sin^{-1}\left(\frac{2\pi}{n_0}\right) \approx \frac{2\pi}{\kappa n_0},
\end{equation}
 and the phase coherence time, over which $|G^{(1)}(0,t)|$ 
decays is 
\begin{equation}
\label{tcoh}
t_{\text{coh}}=\frac{1}{\kappa}\cos^{-1}\left(1-\frac{1}{2n_0}\right) \approx \frac{1}{\kappa\sqrt{n_0}}
.\end{equation}
 The first quantum revival occurs at 
\begin{equation}
t_{\text{revival}} = \frac{2\pi}{\kappa}
.\end{equation}

\subsection{Causes of growth of sampling error}
\label{SIMTIMEDIV}
From (\ref{formalpp}), $\log n(t)$ is Gaussian distributed, while $\log\alpha(t)$ is also Gaussian-like, at least at short times when the effect of the nonlinear term is negligible. The variables $\alpha$ and $\beta=n/\alpha$ appearing in observable calculations behave approximately as exponentials of Gaussian random variables, at least over short times. Hence, the properties of 
this kind of random quantity can tell us much about the behaviour of the simulation. 

 If $\xi$ is a Gaussian random variable of mean zero, variance unity, let us define
\begin{equation}
v = v_0e^{\sigma\xi}
\end{equation}
with real positive $\sigma$. The even $m$ moments of $\xi$ are 
\begin{equation}\label{ximoments}
\lim_{\mc{S}\to\infty}\langle\xi(t)^m\rangle_{\rm s}=\frac{m!}{2^{\,m/2}(m/2)!},
\end{equation}
 while the odd $m$ moments are zero. This can be used to obtain
\begin{equation}
\lim_{\mc{S}\to\infty}\langle v\rangle_{\rm s} = \ba{v} = v_0\exp\left(\frac{\sigma^2}{2}\right)
\end{equation}
and
\begin{equation}\label{varvsigma}
\lim_{\mc{S}\to\infty}\var{\frac{v}{\ba{v}}} = e^{\sigma^2}-1.
\end{equation}
I.e. the standard deviation of $v$ is linear in $\sigma$ for $\sigma\lesssim1$, but grows  rapidly once $\sigma\gtrsim 1$.

\begin{figure}
\vspace*{6pt}\center{\includegraphics[width=300pt]{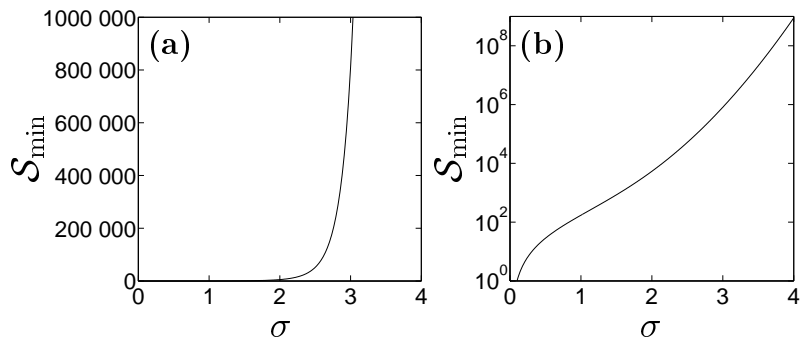}}\vspace*{-12pt}
\caption{\label{FIGvargaus} Number of samples $\mc{S}_{\rm min}$ required to obtain a single significant digit of precision in $\ba{v}$, the estimate of the mean of the random variable $v=e^{\sigma\xi}$. }
\end{figure}

With finite samples $\mc{S}$, the fluctuations of $\langle v \rangle_{\rm s}$ around $\ba{v}$ are usually estimated 
to be 
\begin{equation}
\Delta \ba{v} \approx \sqrt{\frac{\var{v}}{\mc{S}}}
\end{equation}
using the Central Limit Theorem (CLT). Hence, the number of samples needed to achieve a given relative precision is 
\begin{equation}
\mc{S}\propto\var{v/\ba{v}}.
\end{equation}
This is shown in Figure~\ref{FIGvargaus}.
In particular, we see that simulations with reasonable numbers of trajectories $\mc{S}\lesssim\order(10^6)$ only give useful precision while 
\begin{equation}\label{sigma10}
\sigma^2\lesssim 10.
\end{equation}

In actual calculations of moments, from \eqref{moments} we see that common observables are estimated by averages $\langle f\rangle_{\rm s}$, where $f$ is polynomial in $\alpha$ and/or $n$. Because the exact solutions (\ref{formalpp}) are Gaussian-like,  $f$ behaves similarly to the idealized variable $v$ discussed above, and useful precision will be limited by 
\begin{equation}\label{flimit}
\var{\log|f|}\lesssim \order(10).
\end{equation}
in analogy with \eqref{sigma10}.
We have switched to $\order(10)$ rather than just $10$ here, because the details of the distribution of $\log|f|$ may differ from a Gaussian at long times due to the nonlinear term $\propto\int n$ in \eqref{formalpp}.

Observables that are first order in $\op{a}$, $\dagop{a}$ (such as $G^{(1)}(0,t)$), or first order in $\dagop{a}\op{a}$ will be limited by $\var{\,\log|f|\,}\approx\var{\,\log|\alpha(t)|\,}$ or $\var{\,\log|n(t)|\,}$, respectively.

\subsection{Logarithmic variances}
\label{PPLOGVARS}
The time evolution of the variances of the logarithmic variables can be calculated in the limit $\mc{S}\to\infty$ from the formal solutions (\ref{formalpp}). For the phase-dependent variables it is more convenient to consider just the mean of the variances of the two complementary variables $\alpha$ and $\beta$:
\begin{equation}\label{mcVdef}
\mc{V} = \frac{1}{2}\left\{\var{\,\log|\alpha(t)|\,} + \var{\,\log|\beta(t)|\,}\right\}.
\end{equation}
Consider coherent-state starting conditions $\op{\rho}(0)=\op{\Lambda}_0/\text{Tr}[\op{\Lambda}_0]$ such that $n(0)=n_0$  for all trajectories.
We will use the notation
\begin{equation}\label{niio}
n_0=n'_0+in''_0,
\end{equation}
etc. for real and imaginary parts.

 Noting that 1): $\log n$ is exactly Gaussian distributed, and 2): in the Ito calculus noises are uncorrelated with any variables at the same or previous times, one obtains from (\ref{formalpp}):
\begin{equation}
\mc{V} = \kappa t/2 +\kappa^2\int_0^t ds \int_0^t ds'\langle n''(s)n''(s')\rangle_{\rm s}.
\end{equation}
Now if $s'>s$, then from (\ref{zetadef}) $\zeta_j(s') = \zeta_j(s)+$\mbox{$\widetilde{\zeta}_j(s'-s)$}, where $\widetilde{\zeta}_j(t)$ are independent of the $\zeta_j$ but again with $\langle\wt{\zeta}_j(t)\wt{\zeta}_k(s)\rangle_{\rm s}=\delta_{jk}\text{min}[t,s]$.
It is convenient to define $\zeta^{\pm} = [\zeta_1\pm\zeta_2]/\sqrt{2}$. Then for $s'>s$
\begin{eqnarray}\fl\eqalign{
\langle n''(s) n''(s') \rangle_{\rm s} 
= e^{-\gamma(s+s')}\langle e^{-2\sqrt{\kappa}\zeta^-(s)}\rangle_{\rm s}\langle e^{-\sqrt{\kappa}\widetilde{\zeta}^-(s'-s)}\rangle_{\rm s}\ \langle\cos[\sqrt{\kappa}\widetilde{\zeta}^+(s'-s)]\rangle_{\rm s}\\
\qquad\times\left\{ (n'_0)^2\langle\sin^2[\sqrt{\kappa}\zeta^+(s)]\rangle_{\rm s}
+(n''_0)^2\langle\cos^2[\sqrt{\kappa}\zeta^+(s)]\rangle_{\rm s} 
+ n'_0n''_0\langle\sin[2\sqrt{\kappa}\zeta^+(s)]\rangle_{\rm s}\right\}.}
\end{eqnarray}
The trigonometric averages can be evaluated using (\ref{ximoments}). After some algebra and integration, one obtains
\begin{subequations}\begin{eqnarray}\label{vpp}\fl\eqalign{
\lim_{\mc{S}\to\infty}\mc{V} =&\ \kappa t/2 -\kappa^2n'_0\left\{\frac{1-e^{-\gamma t}(1+\gamma t)}{\gamma^2}\right\} \\
&+\kappa^2|n_0|^2\left\{\frac{1}{q-\gamma}\left[\frac{1-e^{-\gamma t}}{\gamma}+\frac{e^{-qt}-1}{q}\right]-\frac{1}{2}\left(\frac{1-e^{-\gamma t}}{\gamma}\right)^2\right\},
}\end{eqnarray} 
where 
\begin{equation}
q=2(\gamma-\kappa)
\end{equation}
 is a ``damping strength'' parameter.
 Also, 
\begin{equation}\label{nvarpp}
\var{\log|n(t)|} = \kappa t.
\end{equation}\end{subequations}

\subsection{Special cases}
\label{SPECPP}
Two cases are of particular interest:
With \textbf{no damping} $\gamma=0$, $\mc{V}$ becomes 
\begin{equation}\label{gzeroV}
\mc{V} = \kappa t/2 - (\kappa t)^2n'_0/2 +\frac{1}{3}(\kappa t)^3 |n_0|^2 + \order[(\kappa t)^4|n_0|^2]
\end{equation}
at low times $2\kappa t\ll 1$, and 
\begin{equation}\label{gzeroVbigt}
\mc{V} \approx \kappa t/2 + \frac{1}{4}|n_0|^2e^{2\kappa t}
\end{equation}
at $t\gg1/2\kappa$.

If the damping is small  but nonzero ($\gamma t\ll 1$),  (\ref{gzeroV}) becomes modified to 
\begin{equation}
\mc{V}\approx\kappa t /2 - n'_0(\kappa t)^2[1-\frac{2}{3}\gamma t]/2 +\frac{1}{3}|n_0|^2(\kappa t)^3[1-\frac{5}{4}\gamma t]
\end{equation}
 to order $(\gamma t)^2$ and $(\kappa t)^4$. Also, while $q<0$, (\ref{gzeroVbigt}) becomes 
$\mc{V}\approx\kappa t/2 +\kappa^2|n_0|^2e^{-q t}/q(q-\gamma)$.

With \textbf{strong damping} when  $q>0$, the long time (i.e. $qt\gg1$) behaviour becomes asymptotic to a simple
linear increase with time
\begin{equation}\label{ppvardamp}
\mc{V} = \kappa t/2 - b,
\end{equation}
where
\begin{equation}\label{bdef}
b=\frac{\epsilon^2}{2}[2n'_0-|n_0|^2\epsilon/(1-\epsilon)]
\end{equation}
 is a constant, and $\epsilon=\kappa/\gamma\in[0,1]$. We note
$b$ is positive for strong damping, but later tends towards $-\infty$ as $\epsilon\to 1$. The minimum damping required for positive $b$ grows with mode occupation $n_0$.
The short time behaviour when $\gamma t\ll 1$ is of the same form as the weakly-damped case above.

Referring back to the condition (\ref{flimit}), we see that while the $n(t)$ distribution remains usable for a relatively long time ($t\lesssim \order(10/\kappa))\approx 1.5 t_{\rm revival}$, the phase-dependent variables can rapidly acquire very broad distributions (as per \eqref{gzeroVbigt}) when damping is small and mode occupation is large. \textbf{This is what limits simulation usefulness.}

\subsection{Useful simulation time: analytic expressions}
\label{SIMTIMEANALYTIC}

While moments containing only the mode occupation $\op{n}=\dagop{a}\op{a}$ can be evaluated for long times of order $t_{\rm revival}$, this is due to special symmetries present only in the single-mode system. With inter-mode coupling,  large spreads in any single $\alpha_{\bo{n}}$ will rapidly feed into the remainder of the variables via the coupling terms $\propto \omega_{\bo{nm}}$ in the equations (\ref{itoequations}). In particular, evolution of ``occupation'' variables $n_{\bo{n}}$ is non-trivially coupled to the ``phase'' variables $\alpha_{\bo{n}}$.
With the $M$-mode systems in mind, then, we are primarily interested in precision of observable estimates in the single-mode model that involve \textit{phase} variables $\alpha$ or $\beta$, not just $n$ (e.g. $G^{(1)}$).

Using the limit (\ref{flimit}) and (\ref{vpp}) or its special cases in Section~\ref{SPECPP}, useful simulation times $t_{\rm sim}$ can be estimated for coherent-state initial conditions (\ref{offkernel}). For non-coherent-state initial distributions each separate trajectory still starts out as a sample of an off-diagonal  coherent-state kernel (\ref{offkernel}). The useful simulation time is then determined by the shortest useful simulation time to be expected from any of the $\mc{S}$ samples. 

For \textbf{large mode occupation with no damping} (the worst case), the term in $|n_0|^2(\kappa t)^3$ dominates $\mc{V}$ at short times $\kappa t \ll 1$, and 
\begin{equation}\label{tsim1modepp}
t_{\rm sim} \approx \frac{\order(3)}{\kappa|n_0|^{2/3}}.
\end{equation}
Checking back, $\kappa t\approx\order(3)/|n_0|^{2/3}$, so (\ref{tsim1modepp}) is consistent with the short time assumption for $|n_0|\gg \order(5)$. 
At weak nonzero damping $\gamma t\ll 1$, (\ref{tsim1modepp}) becomes modified to 
\begin{equation}
t_{\rm sim}\approx \order(3)/\kappa|n_0|^{2/3}[1+\order(\gamma/\kappa|n_0|^{2/3})].
\end{equation}

At \textbf{small occupations} $n_0\ll1$, simulation times are longer, and the regime where(\ref{gzeroVbigt}) applies is reached. For a significant range of $|n_0|\ll1$, the  term $\propto|n_0|^2$ dominates $\mc{V}$, and
the useful simulation time scales very slowly with $|n_0|$ as 
\begin{equation}\label{midtsimpp}
t_{\rm sim} \approx \frac{\order(1)-\frac{1}{2}\log|n_0|}{-q/4}.
\end{equation}
With weak damping, $-q/2\to\kappa$.  At even smaller $n_0$, the $\kappa t$ term in $\mc{V}$ is greater, and 
\begin{equation}\label{lowtsimpp}
t_{\rm sim} \approx \order(10)/\kappa.
\end{equation}

For sufficiently \textbf{strong damping} such that $q>0$, and $qt_{\rm sim}\gg 1$, (\ref{ppvardamp}) applies, and long simulation times are possible:
\begin{equation}\label{tsimgamma}
t_{\rm sim} \approx \order(b+10)/\kappa.
\end{equation}

\section{Empirical simulation time results}
\label{PPEMPIRICAL}
\begin{figure}\begin{center} 
{\includegraphics[width=250pt]{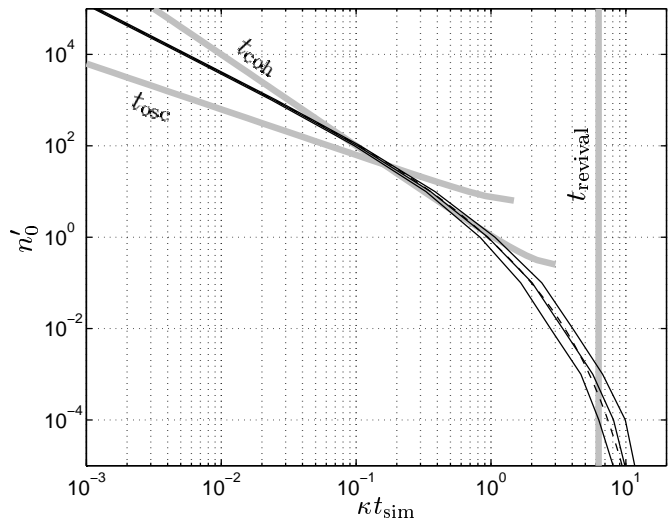}}\vspace*{-12pt}
\end{center} 
\caption{\label{FIGppsim} Useful simulation time $t_{\rm sim}$ for a positive P simulation of undamped one-mode system (\ref{1modeH}) (\textit{solid lines}) Triple lines indicate observed ranges. Dashed line indicates best estimate with (\ref{timefit}). 
}
\end{figure}

To verify the analytic estimates, 
simulations of an undamped single-mode anharmonic oscillator were performed for a wide range of initial coherent states $n_0=n'_0$ from $10^{-5}$ to $10^{10}$. The primary aim was to determine the actual dependence of useful simulation time $t_{\rm sim}$ on mode occupation, particularly for intermediate $n_0$  values $\lesssim\order(1)$ where the simple expression (\ref{tsim1modepp}) does not apply. This is shown in Figure~\ref{FIGppsim}
These results will be useful for subsequent many-mode simulation time estimates in Section~\ref{MMODE}. 

In what follows, the term \textit{useful precision} for an observable $\op{O}$ has been taken to indicate the situation where the estimate $\ba{O}=\langle f\rangle_{\rm s}$ of  $\langle\op{O}\rangle$ using $\mc{S}=10^6$ trajectories has a relative precision of at least $10\%$ at the one sigma confidence level. This is assuming the CLT holds so that 
\begin{equation}
\Delta \ba{O} \approx \sqrt{\frac{\var{f}}{\mc{S}}}
\end{equation}
is used to assess uncertainty in $\ba{O}$. For the model here, we consider useful precision in the magnitude of phase-dependent correlations $|G^{(1)}(0,t)|$, which is the low-order observable most sensitive to the numerical instabilities in the equations. 

Uncertainties in the calculated $t_{\rm sim}$  times arise because the $\Delta|G^{(1)}|$ were themselves estimated from  finite ensembles of $\mc{S}=10^4$ trajectories. The range of $t_{\rm sim}$ indicated in Figure~\ref{FIGppsim} was obtained from 10 independent runs with identical parameters.

Taking the analytic scalings (\ref{tsim1modepp}) and (\ref{midtsimpp}) at high and low  $n_0$ into account, parameters in an approximate curve 
\begin{equation}\label{timefit}
  t_{\text{est}} = \frac{1}{\kappa}\left\{ \left[ {c_1} (n'_0){}^{-2/3} \right]^{-{c_2}} + \left[ 2\log\left(\frac{e^{c_3}}{n'_0{}^{c_4}}+1\right)\right]^{-{c_2}}\right\}^{-1/{c_2}}
\end{equation}
have been fitted to the empirical data. 
Best values of $c_j$ are given in Table~\ref{TABLEtsimfit}, and  in Figure~\ref{FIGppsim} this fit is compared to the empirical data, and seen to be quite accurate.
 $c_1$ characterizes the pre-factor for high $n'_0$, $c_3$ a constant residual $t_{\rm sim}$ at near vacuum, $c_4$ characterizes the curvature at small $n'_0$, while $c_2$ is related to the stiffness of the transition.
 The expression 
(\ref{timefit}) reduces to $c_1n'_0{}^{-2/3}/\kappa$ and $({c_3}-c_4\log n'_0)/(\kappa/2)$, when $n'_0\gg1$ and $n'_0\ll1$, respectively.
Uncertainty $\Delta c_j$ in parameters $c_j$ was worked out by requiring $\sum_{n_0}\{[t_{\text{est}}(c_j\pm\Delta c_j,n_0)-t_{\text{sim}}(n_0)]/\Delta t_{\text{sim}}\}^2 = \sum_{n_0}\{1+([t_{\text{est}}(c_j,n_0)-t_{\text{sim}}(n_0)]/\Delta t_{\text{sim}})^2\}$.

\newcommand{\mm}[2]{{}^{#1}_{#2}}
\begin{table}
\caption{ \label{TABLEtsimfit}
Empirical fitting parameters for maximum useful simulation time $t_{\text{sim}}$ in one mode. The fit is to expression (\ref{timefit}).
}
\begin{indented}\item[]\begin{tabular}{@{}r@{$\ \pm\ $}lr@{$\ \pm\ $}lr@{$\ \pm\ $}lr@{$\ \pm\ $}l}\br
\multicolumn{2}{c}{$c_1$}&\multicolumn{2}{c}{$c_2$}&\multicolumn{2}{c}{$c_3$}&\multicolumn{2}{c}{$c_4$}\\\mr
$2.54$	& $0.16$	& $3.2$		&$\mm{\infty}{1.2}$	& $-0.5$	& $0.3$		& $0.45$	& $0.07$\\
\br
\end{tabular}\end{indented}
\end{table}

\section{Multi-mode simulation times}
\label{MMODE}
  
\subsection{Extrapolation to many modes}
\label{EXTRAP}
Firstly, let us note that for each mode $\bo{n}$ the nonlinearity that leads to rapid growth of distribution broadness at $\mc{V}\approx 10$ depends only on the local variables $\alpha_{\bo{n}}$ and $\beta_{\bo{n}}$. This suggests that extrapolating the single-mode results of the previous section may lead to usable simulation time estimates for the coupled $M$-mode system.

If the distribution of an amplitude variable in any mode acquires broad and/or rapidly growing tails (e.g. due to the single-mode instability caused by the nonlinearity), the coupling terms $\propto\omega_{\bo{nm}}$ rapidly disseminate its influence to the remaining modes, and all observables acquire large uncertainty. This effectively terminates the useful part of the simulation. 
From (\ref{tsim1modepp}) and/or  Figure~\ref{FIGppsim}, we can see that for a single weakly-damped mode $t_{\rm sim}$ drops off rapidly with increasing mode occupation $n_0$. At the same time the distributions of its phase variables broaden rapidly, and this broadness is distributed into the rest of the modes. Thus, the most highly occupied mode of a many-mode system limits the overall simulation time. 

\subsection{Simulation time estimate}
\label{ESTIM}
If we assume (for the time being) that inter-mode coupling does not significantly affect the time in which the most highly occupied mode reaches $\mc{V}\approx 10$, one estimates that 
\begin{equation}\label{ppsimtest}
t_{\rm sim} \approx t_{\rm est}(\text{max}_{(t\lesssim t_{\rm est},\bo{n})}[\ba{n}_{\bo{n}}(t)]),
\end{equation}
where $\ba{n}_{\bo{n}}(t)$ is the mean occupation $\langle\dagop{a}_{\bo{n}}\op{a}_{\bo{n}}\rangle$ at time $t$, and $t_{\rm est}$ is given by (\ref{timefit}). 
Using the mean value $\ba{n}$ appears justified since for usefully precise simulations at higher occupations we expect $|n(t)-\ba{n}(t)|\ll\ba{n}(t)$.
For strongly-damped systems (\ref{tsimgamma}) applies rather than $t_{\rm est}$.

For a model arising as the discretization of a continuum field model (\ref{model}) with density $\rho(\bo{x})$, $\ba{n}_{\bo{n}}\approx\rho(\bo{x}_{\bo{n}})\Delta V$. When $\text{max}(\ba{n}_{\bo{n}}(t))\gg1$, the estimated simulation time becomes 
\begin{equation}\label{ppsimtime}
t_{\rm sim} \approx\left(\frac{2.5\hbar}{g}\right)\frac{(\Delta V)^{1/3}}{(\max_{(t\lesssim t_{\rm est},\bo{x})}[\rho(\bo{x})])^{2/3}}.
\end{equation}

What was perhaps not obvious before obtaining the above expression is that \textbf{coarser lattices lead to longer simulations}. The useful time scales as $(\Delta V)^{1/3}$ in the strong interaction limit. Thus, when simulating a field model, it is worthwhile to use the coarsest lattice for which physical predictions are not affected.

\subsection{Influence of kinetics in the multi-mode case}
\label{PPKINETIC}
  Linear coupling between modes is what complicates the physical behaviour, and consequently also the noise behaviour in a simulation.  Let us use the label $z_{\bo{n}}$ for either $\alpha_{\bo{n}}$ or $\beta_{\bo{n}}$, and look at the short time fluctuations of the phase-space variables. These fluctuations can be separated as
\begin{equation}\label{zfluct}
\delta z_{\bo{n}}(t) \approx \delta^{\rm direct} z_{\bo{n}}(t) + \delta^{\rm kinetic} z_{\bo{n}}(t)
,\end{equation}
into the ``direct'' part due to local interparticle scattering as considered in detail for the single-mode anharmonic oscillator (\ref{1modeH}), and secondary ``kinetic'' fluctuations. These are due mostly to transfer of direct fluctuations between modes by the kinetic inter-mode coupling, at least initially. 

If $\delta^{\rm direct} z_{\bo{n}}(t) \gg \delta^{\rm kinetic} z_{\bo{n}}(t)$, then properties based on the single-mode analysis (such as the simulation time estimates (\ref{ppsimtest}) and (\ref{ppsimtime})) will be qualitatively accurate.
A rough calculation for uniform gases at density $\rho$ (see Appendix~\ref{APPKINETIC} for details) shows that at short times,
\begin{equation}\label{varikd}
\frac{
\var{\,|\delta^{\rm kinetic} z_{\bo{n}}(t)|\,}}{
\var{\,|\delta^{\rm direct} z_{\bo{n}}(t)|\,}}\approx  \order\left[\frac{\hbar^2t^2\pi^4}{60m^2(\Delta V)^{4/D}}\right]
,\end{equation}
while the right hand side is $\ll 1$.
This is independent of  the uniform density itself, and indicates
that the additional fluctuations due to kinetic effects: 
\begin{enumerate}
\item
 Become relatively more important with time.
\item Are more dominant for fine lattices.
\end{enumerate}
Once (\ref{varikd}) becomes $\gtrsim\order(1)$, deviations from the simulation time estimates of Section~\ref{ESTIM} can be expected, although this simple analysis does not tell us  whether inter-mode coupling decreases or increases simulation time. In general a decrease seems intuitively more likely, but an increase may occur if $\delta^{\rm direct} z_{\bo{n}}(t)$ and $\delta^{\rm kinetic} z_{\bo{n}}(t)$
become appropriately correlated to reduce overall fluctuations.

\section{Multi-mode example: 1D gas}
\label{MEX}

\subsection{Model}

As an example of many-mode dynamical simulations, we have simulated  a uniform one-dimensional gas of bosons with density $\rho$ and inter-particle \mbox{$s$-wave} scattering length $a_s$. The lattice is chosen with a spacing $\Delta{\rm x}_d\gg a_s$ so that interparticle interactions are effectively local at each lattice point, and the lattice Hamiltonian \eqref{Hamiltonian} applies. The stochastic equations to simulate are \eqref{itoequations}. Periodic boundary conditions are assumed.

The initial state is taken to be a $T=0$ coherent wavefunction, which is a stationary state of the ideal gas (i.e. when  $a_s=g=0$). Subsequent evolution is with constant $a_s>0$, so that there is a disturbance at $t=0$ when interparticle interactions are rapidly turned on.

Physically, this would correspond to the disturbance created in a BEC by rapidly increasing the scattering length at $t\approx 0$ by tuning the external magnetic field near a Feshbach resonance. Behaviour found in this uniform system will carry over without qualitative change to  a 1D BEC confined on a length significantly larger than the length scale of observed dynamical phenomena. Most of these phenomena extend over  the order of a healing length
\begin{equation}\label{heallength}
\xi^{\rm heal} = \frac{\hbar}{\sqrt{2m\rho g}}
.\end{equation}
 This is the minimum length scale over which a local density inhomogeneity  in a Bose condensate wavefunction can be in balance with the quantum pressure due to kinetic effects\cite{heallength}.

A useful time scale here is 
\begin{equation}\label{txidef}
t_{\xi} = \frac{m(\xi^{\rm heal})^2}{\hbar} = \frac{\hbar}{2\rho g}
\end{equation}
(the ``healing time''), which is approximately the time needed for the short-distance $\order(\xi^{\rm heal})$ inter-atomic correlations to equilibrate after the disturbance\cite{inprep}.

\subsection{Spatial correlations}
	Here we calculate the dynamics of spatial correlation functions, which has not been previously investigated from first principles in this model. We consider values averaged over the location of one particle out of the pair/triplet since this is a uniform gas:

\textbf{First order}:
\begin{equation}
\ba{g}^{(1)}(\bo{x}_{\bo{n}}) = \frac{1}{M}\sum_{\bo{m}}
\frac{\langle\dagop{a}_{\bo{m}}\op{a}_{\bo{m+n}}\rangle}{\sqrt{\langle\dagop{a}_{\bo{m}}\op{a}_{\bo{m}}\rangle\langle\dagop{a}_{\bo{m+n}}\op{a}_{\bo{m+n}}\rangle}}.
\end{equation}
This is a phase-dependent correlation function. Its magnitude $|\ba{g}^{(1)}|$ tells one the degree of first-order spatial coherence, while its phase gives the relative phase of the wavefunction at distances $\bo{x}_{\bo{n}}$.

\textbf{Second order:}
\begin{equation}
\ba{g}^{(2)}(\bo{x}_{\bo{n}}) = \frac{1}{M}\sum_{\bo{m}}
\frac{\langle\dagop{a}_{\bo{m}}\dagop{a}_{\bo{m+n}}\op{a}_{\bo{m}}\op{a}_{\bo{m+n}}\rangle}{\langle\dagop{a}_{\bo{m}}\op{a}_{\bo{m}}\rangle\langle\dagop{a}_{\bo{m+n}}\op{a}_{\bo{m+n}}\rangle}.
\end{equation}
This describes the likelihood of finding two particles at a distance $\bo{x}_{\bo{n}}$ from each other, relative to what is expected of a coherent field. For a bunched field, $\ba{g}^{(2)}(\bo{x})>1$ (e.g. a thermal state has $\ba{g}^{(2)}(\bo{x})=2$), while spatial antibunching is evidenced by $\ba{g}^{(2)}(\bo{x})<1$.

\textbf{Third order:}
\begin{equation}
\ba{g}^{(3)}(\bo{x}_{\bo{n}}) = \frac{1}{M}\sum_{\bo{m}}\frac{\langle\dagop{a}_{\bo{m}}\dagop{a}_{\bo{m+n}}\dagop{a}_{\bo{m-n}}\op{a}_{\bo{m}}\op{a}_{\bo{m+n}}\op{a}_{\bo{m-n}}\rangle}{\langle\dagop{a}_{\bo{m}}\op{a}_{\bo{m}}\rangle\langle\dagop{a}_{\bo{m+n}}\op{a}_{\bo{m+n}}\rangle\langle\dagop{a}_{\bo{m-n}}\op{a}_{\bo{m-n}}\rangle}.
\end{equation}
  This three-particle correlation function describes the likelihood of three particles with (equal) distances $\bo{x}_{\bo{n}}$ between nearest neighbours (relative to a coherent field). In a BEC, the rate of three-body recombination, which can limit the condensate's lifetime, is proportional to $\ba{g}^{(3)}(0)$\cite{Kagan-88}.

\subsection{Comparison to simulation time estimates}

A number of positive P simulations of a one-dimensional gas were made, with densities of $\rho=100/\xi^{\rm heal}$,$10/\xi^{\rm heal}$, and $1/\xi^{\rm heal}$, and various lattice spacings. The number of lattice points was varied between simulations taking on values from $M=25$ to $M=500$ depending on circumstances to encompass and resolve the phenomena seen. More  lattice points were required at the higher densities where correlations were weaker, resulting in longer  simulation times overall.

\begin{figure}
\vspace*{6pt}\center{\includegraphics[width=200pt]{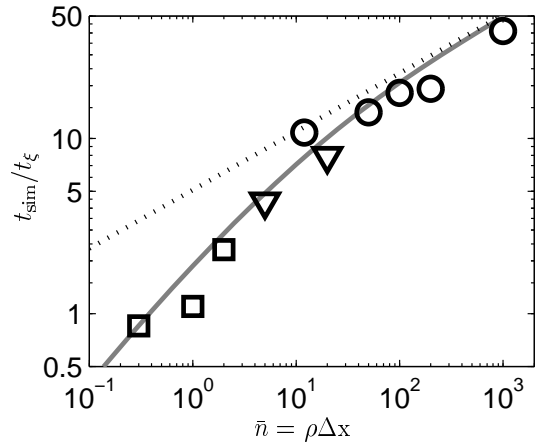}}\vspace*{-10pt}
\caption{\label{FIGMtsim}
Comparison of useful simulation time $t_{\rm sim}$ to the estimates \eqref{ppsimtime} (dotted) and \eqref{ppsimtest} (grey line). $t_{\rm sim}$ based on precision in $\ba{g}^{(2)}(x)$.
Data points  for simulations of 1D gases of densities $100/\xi^{\rm heal}$, $10/\xi^{\rm heal}$, and $1/\xi^{\rm heal}$, are shown with {\scshape circles, triangles,} and {\scshape squares}, respectively. 
 $\mc{S}=10^4$ trajectories.
}
\end{figure}

Figure~\ref{FIGMtsim} compares the actual useful simulation times in simulations  with what was predicted by expressions \eqref{ppsimtest} and \eqref{ppsimtime}.  We can see that the fit is remarkably good despite inter-mode coupling.
It is clearly seen also that the simulation time decreases with $\Delta{\rm x}=\ba{n}/\rho$.

Note that values of $t_{\rm sim}$  are based on precision in $\ba{g}^{(2)}$. Precision tends to become worse as higher-order moments of $\op{a}$ or $\dagop{a}$ become needed, in e.g. $\ba{g}^{(3)}$. 
This is simply because higher-order moments require the averaged quantity $f$ to be a higher-order polynomial in $\alpha$ and $\beta$, and in effect $\var{\log|f|}\approx c^2\mc{V}$, with $c$ being approximately the order of moment.

\begin{figure}
\vspace*{6pt}\center{\includegraphics[width=300pt]{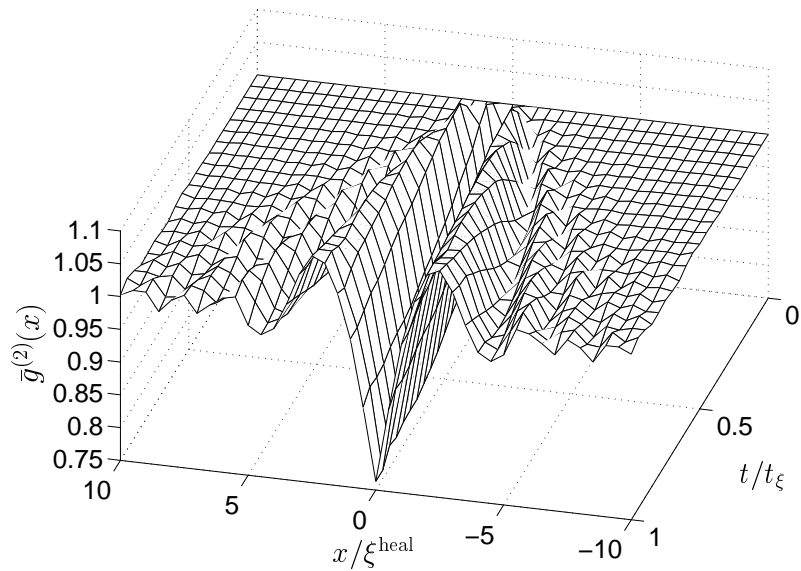}}\vspace*{-10pt}
\caption{\label{FIGMg2}
Second-order correlations in a uniform 1D gas with density $\rho=1/\xi^{\rm heal}$. 50 lattice points, $\Delta {\rm x}=\xi^{\rm heal}/2$, and $\mc{S}=10^4$ trajectories.}
\end{figure}

\begin{figure}
\vspace*{6pt}\center{\includegraphics[width=300pt]{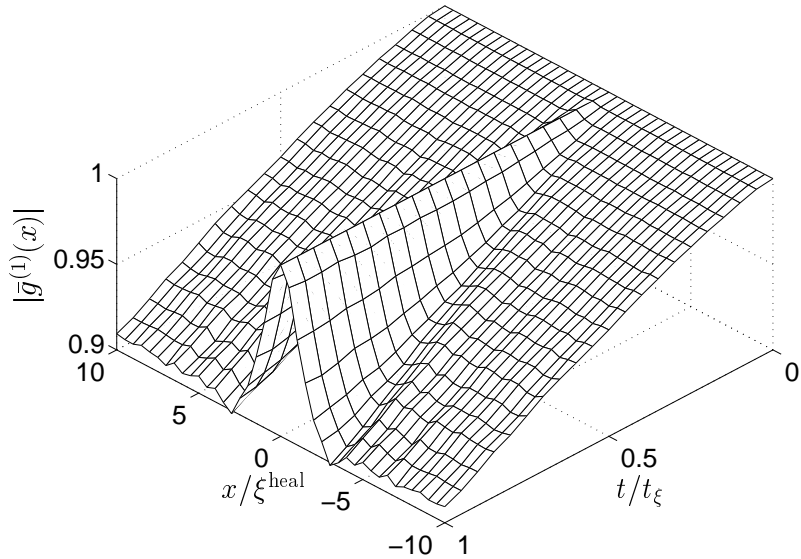}}\vspace*{-10pt}
\caption{\label{FIGMg1}
First-order correlations. Details as in Figure~\ref{FIGMg2}.}
\end{figure}

\begin{figure}
\vspace*{6pt}\center{\includegraphics[width=300pt]{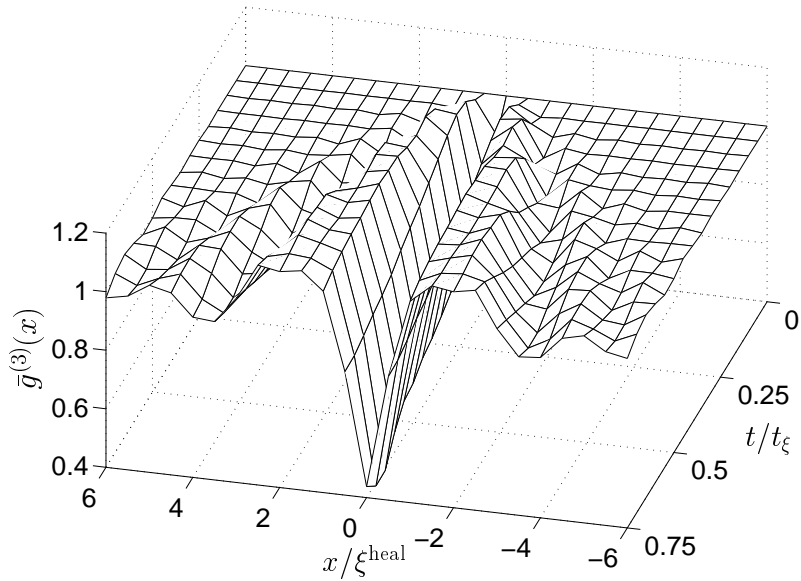}}\vspace*{-10pt}
\caption{\label{FIGMg3}
Third-order correlations. Details as in Figure~\ref{FIGMg2}.}
\end{figure}

\begin{figure}
\vspace*{6pt}\center{\includegraphics[width=250pt]{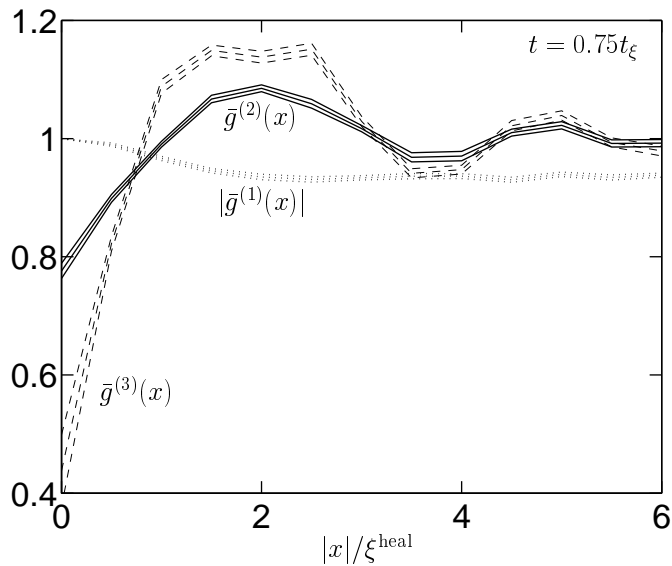}}\vspace*{-10pt}
\caption{\label{FIGMg}
Correlations at $t=0.75t_{\xi}$. Triple lines indicate error bars at one $\sigma$. Details as in Figure~\ref{FIGMg2}.}
\end{figure}

Figures~\ref{FIGMg2} to~\ref{FIGMg} show calculated correlation functions for the example system $\rho=1/\xi^{\rm heal}$. We see enhanced pairing/tripling of particles at several preferred relative distances. These preferred interparticle distances 
increase with time.  
Note that the local density $\rho(\bo{x},t)$ is completely invariant, and \textit{this wave-like behaviour is seen only in the correlations}.
It is interesting to note that the overall disturbance advances at a much faster rate than the peaks/troughs of the correlations. We also see that while long-range coherence is steadily lost, there is a short distance scale corresponding to the distance of the nearest two-particle correlation peak on which phase coherence is enhanced compared to the long-range situation.

Significantly more detail regarding  calculations with the uniform gas and the correlation waves seen can be found in \cite{inprep}, Chapter 10.


\section{Conclusions}
\label{CONCLUSIONS}
We have obtained the following:
\begin{itemize}
\item Simulation time estimates \eqref{ppsimtest} and \eqref{ppsimtime} for many-mode positive P simulations, the former using the fitting curve \eqref{timefit} and empirical data of Table~\ref{TABLEtsimfit}, the latter applicable at maximum mode occupation $\gg1$. It was found that useful simulation time decreases with density and increases with lattice spacing (coarseness).
\item These simulation time estimates were found to be accurate for simulations of uniform 1D gases.
\item Dynamics of  spatial correlation behaviour (1st to 3rd order) in an interacting uniform 1D gas after disturbance due to change in scattering length. Propagating ``correlation waves'' in 2nd- and 3rd-order correlations (Figures~\ref{FIGMg2} to ~\ref{FIGMg}) were seen.
\item Detailed characterization of  the behaviour of a single spatial mode \eqref{1modeH} in an interacting Bose gas model.
\item Estimates \eqref{varikd} of relative importance of direct and coupling-induced fluctuations in phase variables at short times. 
\end{itemize}
A comparison with other more recently proposed methods was made in Table~\ref{TABLEmethods}, and it is seen that each has its own merits, depending on the model. In broad terms the advantages of the positive P /gauge P method are its simplicity and versatility: Equations to simulate are just the GP mean-field equations with added Gaussian noise which, remarkably, is sufficient to account for all quantum mechanical features. Also, there is no truncation of entanglement, no hard-wired conservation laws, and open systems are the standard. The method is trivially adaptable to parallel computation: just run each trajectory separately.

The dynamical timescales that can be simulated in a given system using the positive P method, can be easily assessed 
``on the back of an envelope'' using the simulation time estimates given here.

\appendix
\section{Scattering and kinetic fluctuations}
\label{APPKINETIC}
The aim here is to assess the relative strength of fluctuations in phase-dependent variables due to direct noise terms $\propto\sqrt{\kappa}$ and to mixing of noise ($\propto\omega_{\bo{nm}}$) from other modes by kinetic processes. 
A uniform gas of density $\rho$, with $\gamma=V_{\rm ext}=0$ is considered, and we use the notation of Section~\ref{PPKINETIC}.
We denote the linear coupling terms in the equations (\ref{itoequations}) as  
\begin{eqnarray}
dz_{\bo{n}} &=& -i\sum_{\bo{m}}\omega_{\bo{nm}}z_{\bo{m}}\,dt + \dots \nonumber\\
 &=& -i\mc{E}^{(z)}_{\bo{n}}\,dt + \dots,
\end{eqnarray}
where $z$ can denote either the variable $\alpha$ or $\beta^*$.
 Kinetic terms in the Hamiltonian (\ref{model}) lead to 
\begin{equation}\label{mcEdef}
\mc{E}_{\bo{n}}^{(z)} = \frac{\hbar}{2mM}\sum_{\bo{m},\wt{\bo{n}}} z_{\bo{m}} |\bo{k}_{\wt{\bo{n}}}|^2 e^{i\bo{k}_{\wt{\bo{n}}}\cdot(\bo{x}_{\bo{n}}-\bo{x}_{\bo{m}})}
,\end{equation}
on a lattice with $M$ points.  $\wt{\bo{n}}=(\wt{n}_1,\dots,\wt{n}_D)$ labels lattice points in a Fourier space such that 
$\bo{k}_{\wt{\bo{n}}} = (\Delta{\rm k}_1\wt{n}_1,\dots,\Delta{\rm k}_D\wt{N}_D)$, where $\Delta k_d = 2\pi/L_d$, and $L_d$ is the  full lattice extent in the $d$ dimension.

Since the gas is uniform, mean conditions at each lattice point must be identical, and random variables can be written as the sum of a global mean part and local fluctuations:
\begin{eqnarray}\eqalign{
\mc{E}_{\bo{n}}^{(z)}(t) =& \ba{\mc{E}}^{(z)}(t) + \delta\mc{E}_{\bo{n}}^{(z)}(t) \\
z_{\bo{n}}(t) =& \ba{z}(t) + \delta z_{\bo{n}}(t)
}\end{eqnarray}
Due to lack of external forces  $|\ba{z}(t)|=\sqrt{\rho\Delta V}$.

Let us denote the $z_{\bo{n}}$ fluctuations when no inter-mode coupling is present (i.e. $\mc{E}_{\bo{n}}^{(z)}=0$) as $\delta^{\rm direct}z_{\bo{n}}$. 
In this case the formal solutions (\ref{formalpp}) apply to each mode independently. 
At \textit{short times} $\kappa t\ll 1$ the nonlinear drift term initially acts mainly as a deterministic time-varying phase, and so the dominant fluctuating contribution would be due directly to the noise terms.  Then 
\begin{eqnarray}\eqalign{\label{roughz}
\log \alpha_{\bo{n}}(t) \approx& \log\ba{\alpha}(t) +i\sqrt{i\kappa}\zeta_{1\bo{n}}(t)\\
\log \beta_{\bo{n}}(t) \approx& \log\ba{\beta}(t) +\sqrt{i\kappa}\zeta_{2\bo{n}}(t).
}\end{eqnarray}
with $\zeta_{j\bo{n}}$ being random variables of the  (\ref{zetadef}) kind and independent for each $\bo{n}$.
At short enough times that $\var{\log |z_{\bo{n}}(t)|}\ll 1$ , one then has 
\begin{eqnarray}\label{ddirectdef}
z_{\bo{n}}(t) &\approx& \ba{z}(t) + \ba{z}(t)\left[\log z_{\bo{n}}(t)-\log\ba{z}(t)\right]\nonumber\\
&=& \ba{z}(t) + \delta^{\rm direct}z_{\bo{n}}(t).
\end{eqnarray}
Using (\ref{roughz}), (\ref{ddirectdef}), and the properties (\ref{zetamean}), one finds 
\begin{equation}\label{varidirect}
\var{\,|\delta^{\rm direct} z_{\bo{n}}(t)|\,} \approx \frac{\rho gt}{\hbar}
.\end{equation}

Now for the kinetic-induced fluctuations. At short times while these are small, i.e. $ \var{\,|\delta^{\rm kinetic}|\,} \ll \var{\,|\delta^{\rm direct}|\,}$, the additional fluctuations in the local terms due to the mode-mixing can be ignored, and so the remaining fluctuations due to coupling are just  $\delta^{\rm kinetic} z_{\bo{n}}(t) \approx -i\int_0^t\delta \mc{E}^{(z)}_{\bo{n}}(s)\, ds$. 
Using \eqref{mcEdef}, and substituting in for $z_{\bo{m}}$ with the approximate (direct noise only) short time expression 
$z_{\bo{m}}\approx\ba{z}+\delta^{\rm direct} z_{\bo{m}}$, one obtains
\begin{eqnarray}\fl\eqalign{
\text{var}\left[\,|\delta^{\rm kinetic} z_{\bo{n}}(t)|\,\right]
\approx \left(\frac{\hbar}{2mM}\right)^2
\int_0^tdt' \int_0^t dt''\\
\qquad\times \sum_{\bo{m},\bo{m}',\wt{\bo{n}},\wt{\bo{n}}'}
\langle\delta^{\rm direct} z^*_{\bo{m}'}(t'')\delta^{\rm direct} z_{\bo{m}}(t')\rangle_{\rm s}
 |\bo{k}_{\wt{\bo{n}}}|^2  |\bo{k}_{\wt{\bo{n}}'}|^2 
\ e^{-i\bo{k}_{\wt{\bo{n}}'}\cdot(\bo{x}_{\bo{n}}-\bo{x}_{\bo{m}'})}
e^{i\bo{k}_{\wt{\bo{n}}}\cdot(\bo{x}_{\bo{n}}-\bo{x}_{\bo{m}})}.
}\end{eqnarray}
Since the direct noise at each lattice point in the locally-interacting model is independent, then
$\langle\delta^{\rm direct} z^*_{\bo{m}'}(t'')\delta^{\rm direct} z_{\bo{m}}(t')\rangle_{\rm s}
= \delta_{\bo{m},\bo{m}'} \rho g\,\text{min}[t'',t']/\hbar,$
similarly to (\ref{varidirect}).
After performing the integrations over $t'$ and $t''$ and simplifying the Fourier transforms one obtains
\begin{equation}\label{varvar}
\var{\,|\delta^{\rm kinetic} z_{\bo{n}}(t)|\,}
\approx \var{\,|\delta^{\rm direct} z_{\bo{n}}(t)|\,} \frac{\hbar^2t^2}{12m^2M}\sum_{\wt{\bo{n}}}|\bo{k}_{\wt{\bo{n}}}|^4
.\end{equation}
For a $D$-dimensional system with many modes, one can approximate 
\begin{eqnarray}
\sum_{\wt{\bo{n}}}|\bo{k}_{\wt{\bo{n}}}|^4 &\approx& \frac{M\Delta V}{(2\pi)^{D}}\int  |\bo{k}|^4\,d^{D}\bo{k} =  \frac{c_1(\bo{k}^{\rm max})M\pi^4}{5(\Delta V)^{4/D}}\label{k4int}
,\end{eqnarray}
where $c_1(\bo{k}^{\rm max})$ is a shape factor $\order(1)$ that depends on the ratios between momentum cutoffs $k^{\rm max}_d=\pi/\Delta{\rm x}_d$ in the various lattice dimensions. 
 For example in $1D$, \mbox{$c_1 = 1$}, while for $2D$ and $3D$ when the momentum cutoffs in each dimension are equal, one has $c_1=3\frac{1}{9}$, and $c_1=6\frac{1}{3}$.
Substituting into \eqref{varvar}  one obtains the final estimate (\ref{varikd}).


\section*{References}
\begin{thebibliography}{10}
\bibitem{positiveP1} S.~Chaturvedi, P.~D.~Drummond, and D.~F.~Walls, J.~Phys.~A \textbf{10}, L187-192 (1977). 
\bibitem{positiveP2} P.~D.~Drummond and C.~W.~Gardiner, J.~Phys.~A \textbf{13}, 2353 (1980).
\bibitem{stochasticgauges} P.~D.~Drummond and P.~Deuar, J.~Opt.~B-Quant.~and Semiclass.~Opt. \textbf{5}, S281 (2003).
\bibitem{separabledistributions} M.~Cetinbas and J.~Wilkie, quant-ph/0407036 (2004).
\bibitem{paperB} P.~Deuar and P.~D.~Drummond, submitted to J.~Phys.~A:~Math.~Gen. (2004), cond-mat/0501058. See also \cite{inprep}.
\bibitem{interactionassumptions} A treatment of these issues can be found in A.~J.~Leggett, Rev.~Mod.~Phys. \textbf{73}, 307 (2001). For a more detailed treatment see J.~Weiner, V.~S.~Bagnato, S.~Zilio, and P.~S.~Julienne, Rev.~Mod.~Phys. \textbf{71}, 1 (1999).
\bibitem{removalofboundaryterms} P.~Deuar and P.~D.~Drummond, Phys.~Rev.~A \textbf{66}, 033812 (2002).
\bibitem{inprep} In preparation. See also, P.~Deuar, PhD thesis, The University of Queensland (2004), cond-mat/0507023.
\bibitem{gardiner}
See e.g. C.~W.~Gardiner, \textit{Quantum Noise} (Springer-Verlag, Berlin, Heidelberg, 1991); C.~W.~Gardiner \textit{Handbook of Stochastic Methods} (Springer-Verlag, Berlin New York, 1983). 
\bibitem{kerrsolitons} S.~J.~Carter, P.~D.~Drummond, M.~D.~Reid, and R.~M.~Shelby, Phys.~Rev.~Lett \textbf{58}, 1841 (1987); 
P.~D.~Drummond, R.~M.~Shelby, S.~R.~Friberg, Y.~Yamamoto, Nature \textbf{365}, 307 (1993).
\bibitem{evaporativecooling} P.~D.~Drummond and J.~F.~Corney, Phys.~Rev.~A \textbf{60}, R2661 (1999).
\bibitem{samperrorinbec} M.~J.~Steel, M.~K.~Olsen, L.~I.~Plimak, P.~D.~Drummond, S.~M.~Tan, M.~J.~Collett, D.~F.~Walls, and R.~Graham, Phys.~Rev.~A \textbf{58}, 4824 (1998).
\bibitem{1dbosethermo} P.~D.~Drummond, P.~Deuar, and K.~V.~Kheruntsyan, Phys.~Rev.~Lett. \textbf{92}, 040405 (2004).
\bibitem{carusottodynamix} I.~Carusotto, Y.~Castin, and J.~Dalibard, Phys.~Rev.~A \textbf{63}, 023606 (2001).
\bibitem{carusottothermo} Y.~Castin, and I.~Carusotto, J.~Phys.~B \textbf{34}, 4589 (2001).
\bibitem{DMRG-Caltech} G.~Vidal, Phys.~Rev.~Lett. \textbf{91}, 147902 (2003); G.~Vidal, Phys.~Rev.~Lett. \textbf{93}, 040502 (2004); M.~Zwolak and G.~Vidal Phys.~Rev.~Lett. \textbf{93}, 207205 (2004).
\bibitem{DMRG-Garching} F.~Verstraete, J.~J.~Garcia-Ripoll, and J.~I.~Cirac, Phys.~Rev.~Lett. \textbf{93}, 207204 (2004); F.~Verstraete and J.~I.~Cirac cond-mat/0407066 (2004).
\bibitem{DMRG-bipartiteentanglement} G.~Vidal, J.~I.~Latorre, E.~Rico, and A.~Kitaev, Phys.~Rev.~Lett \textbf{90}, 227902 (2003).
\bibitem{heallength} See e.g. F.~Dalfovo, S.~Giorgini, L.~P.~Pitaevskii, and S.~Stringari, Rev.~Mod.~Phys. \textbf{71}, 463 (1999), p. 481.
\bibitem{Kagan-88} Yu Kagan, B.~V.~Svistunov, and G.~V.~Shlyapnikov, JETP Lett. \textbf{48}, 56 (1988).
\end{thebibliography}
\end{document}